\documentclass[12pt]{article}

\usepackage{graphicx}
\usepackage{color}
\usepackage{mathptmx}      % use Times fonts if available on your TeX system
\usepackage{amssymb}
\usepackage{amsmath}
\usepackage{ifsym}
\usepackage{sidecap}
\usepackage[colorlinks, linkcolor=blue, citecolor=blue, urlcolor=blue]{hyperref}

\usepackage{amssymb}

\begin{document}

\noindent
%\centerline
{\Large On Protecting the Planet Against Cosmic Attack:}

\noindent
%\centerline
{\Large Ultrafast Real-Time Estimate of the Asteroid's}

\noindent
%\centerline
{\Large Radial Velocity\footnote{\sl Preprint submitted to Acta Astronautica}}

\bigskip
\bigskip

\noindent
{\large V. D. Zakharchenko\footnote{\sl E-mail address: zakharchenko\_vd@mail.ru},  I. G. Kovalenko\footnote{\sl E-mail address: ilya.g.kovalenko@gmail.com}
}

\bigskip
\noindent
{\it Volgograd State University,
         Universitetskij Pr., 100,  Volgograd 400062, Russia
}

%\smallskip
%\noindent

\begin{abstract}
A new method for the line-of-sight velocity estimation of a high-speed near-Earth object (asteroid, meteorite) is suggested. The method is based on use of fractional, one-half order derivative of a Doppler signal. The algorithm suggested is much simpler and more economical than the classical one, and it appears preferable for use in orbital weapon systems of threat response. Application of fractional differentiation to quick evaluation of mean frequency location of the reflected Doppler signal is justified. The method allows an assessment of the mean frequency in the time domain without spectral analysis. An algorithm structure for the real-time estimation is presented. The velocity resolution estimates are made for typical asteroids in the X-band. It is shown that the wait time can be shortened by orders of magnitude compared with similar value in the case of a standard spectral processing. \end{abstract}


%%
%% Start line numbering here if you want
%%
%%\linenumbers

\section{Introduction}\label{Introduction}

The eventful month between mid-February and mid-March 2013 reminded the mankind that the asteroid-comet threat is real. `Chebarkul' super-bolide which exploded on February 15, 2013 over the densely populated region of Chelyabinsk caused damage of buildings and injuries of people. Close flybys of two potentially hazardous asteroids 2012 DA14 \cite{DA12} and 2013ET \cite{ET13} at 0.07 and 2.5 lunar distances from Earth, 44 and 100 meters in diameter respectively, both able to ruin a city, also received extensive coverage in mass-media. Besides, the latter one was discovered just six days before its closest approach to Earth while the Chebarkul meteorite was only recorded after it had entered the Earth atmosphere.

These events confirmed a universal truth: with the means of near-Earth space scanning what they are now, a hazardous near-Earth object (NEO) can remain undetected on the distant approaches to Earth, leaving very little time to react. Thus the long discussed problem of protecting the Earth from cosmic threats requires a quick transition to the phase of practical realization.

\begin{figure}
  \includegraphics[width=1\textwidth]{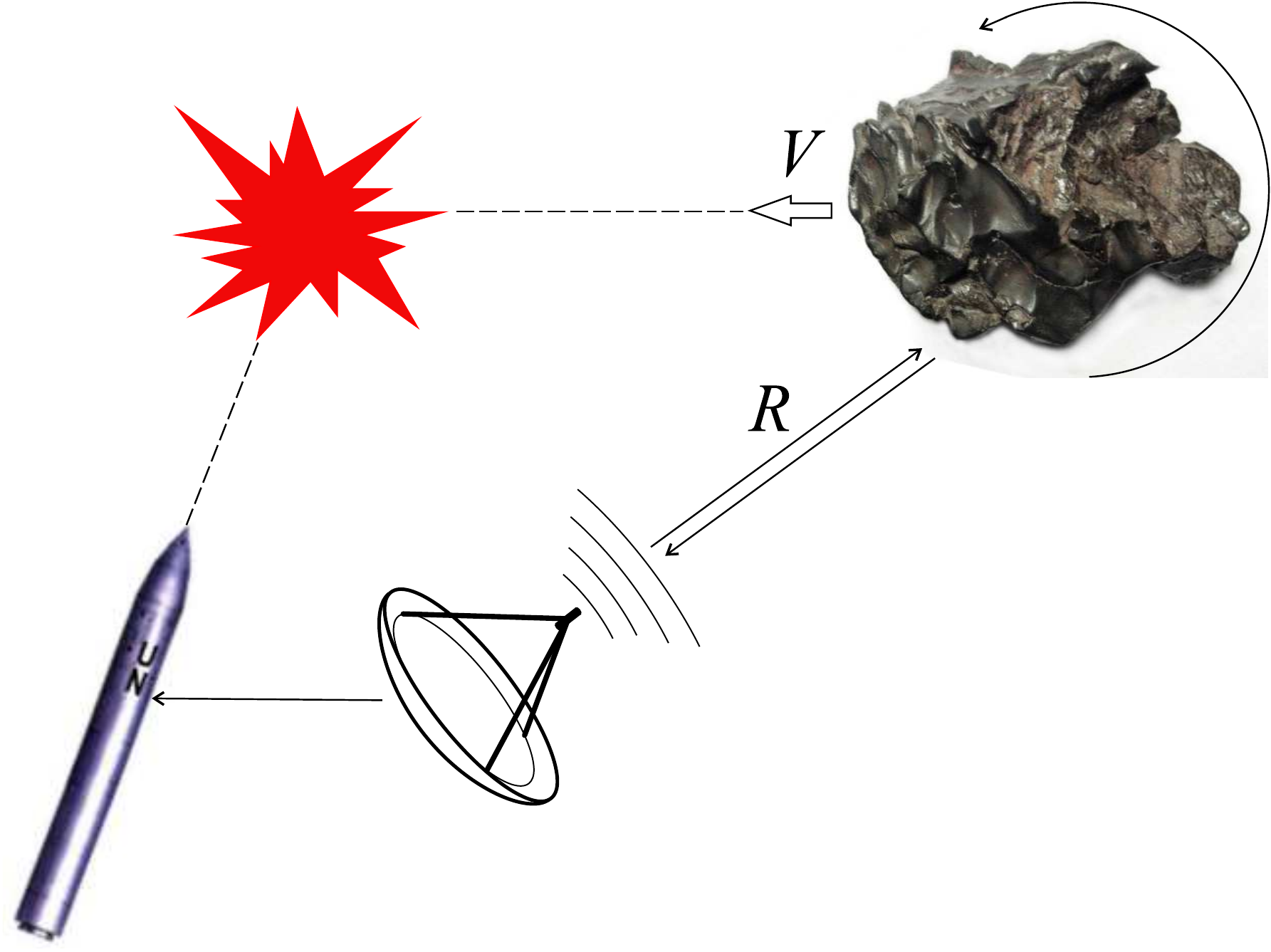}
\caption{A scheme of the asteroid-comet threat counteraction at the final stage.}
\label{fig:1}
\end{figure}

Protecting the planet from cosmic threats means detection of hazardous NEOs, measuring their parameters of motion, trajectory calculation, and their destruction or, at least, prompt alert of the national services for the population protection in the threatened area.

Prior works propose implementing cosmic threat protection through deployment of detection and destruction systems on the near-Earth orbits \cite{Velikhov86, Bolotin, Wie12}. The combat platforms deployment on the geostationary orbit is advantageous for a variety of reasons: (i) ground station contact can be maintained permanently by directional antennae, (ii) an asteroid's approach in the eclipse plane is the most expected; the angle between geostationary orbit plane and eclipse ($23^\circ$) is not large enough to obstruct operation of the communications and targeting facilities.  	

At the most dangerous speeds of $30\div 60$ km/s an asteroid's approach time from the Lunar orbit to the geostationary orbit is about $100\div 200$ minutes. In this scenario the time for location, selection, and maneuver is much limited and the success of action (to eliminate the threat) depends on the precision of aim.

The most reliable and accurate information can be obtained from radiolocation measurements of the asteroid's distance and radial velocity. They allow of robust prediction of motion of the hazardous celestial body \cite{Ostro02,Younger12}. Besides, the radiolocation method is much easier in application in the near-Earth space directly.

When determining the coordinates of the cosmic object one needs a quick, accurate prediction of its trajectory so as to calculate the kill point. It is obvious that at the terminal phase of target approach parameters of the asteroid motion must be measured near-real-time on the weapon system (e.g. missile) when homing (Fig.~\ref{fig:1}). At such large velocity difference between the weapon system and the asteroid ($\gtrsim 30$ km/s against $\sim 10$ km/s) target destruction can be disrupted by a minor inaccuracy. Under the circumstances cost of errors increases manifold and fidelity requirements should be maximized.

The usual trajectory prediction is done by measuring the object's velocity through analyzing the Doppler frequency shift of the reflected radar signal. Measuring Doppler frequency accurately and promptly ensures an effective cosmic shield.

When different points of the object that form the reflected signal are moving at different velocities say, upon the object's rotation, the reflected signal can have a wide spectrum of Doppler frequencies \cite{Ostro97}. which correspond to the spectrum of velocities of reflecting points on the object's surface. In such a case the gravity center of the Doppler signal's\footnote{In radiolocation the Doppler signal means an oscillation achieved as a result of detecting the signal reflected by the target by means of a synchronous detector. In this case we assume that the carrier frequency of the signal emitted by the radar is used as the reference oscillation.}
 energy spectrum is commonly used. This parameter stably corresponds to the center mass motion of the moving object.

In the present paper a the use of original algorithm of NEO's high radial velocity estimate within the time of the Doppler signal arrival is proposed: this algorithm allows an economical use of the timing budget and computational resources of the cosmic shield system when making a trajectory prediction. The authors hope that their contribution to asteroid-comet threats protection will permit, in a small way, a reduction in the probability of hazards to life on the Earth in the foreseeable future.

\section{Evaluation of the mean frequency of the Doppler signal spectrum}\label{Evaluation}

Evaluation of the spectrum gravity center (mean frequency) $\omega_0$  assumes calculation of the energetic spectrum, that is, spectral processing of the signal $x(t)$, which requires bulk memory and, above all, a significant amount of processing time. The latter should be considered unacceptable on tactical grounds with the problem in hand.

\begin{figure}
  \includegraphics[width=1\textwidth]{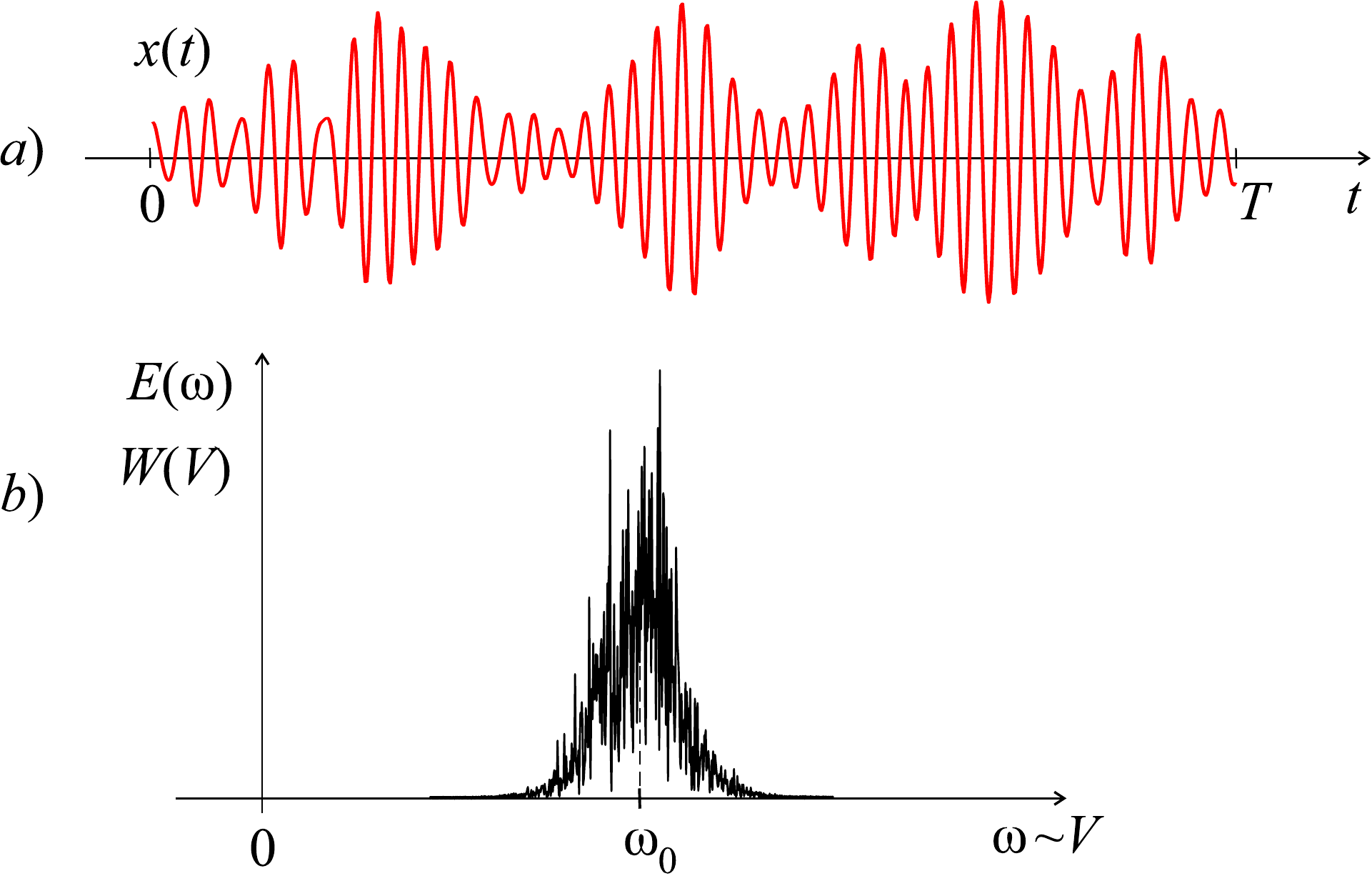}
\caption{The Doppler signal (a) and its energy spectrum (b).}
\label{fig:2}
\end{figure}

The moments method is widely used in the signal theory for evaluation of the frequency spectrum parameters  \cite{Gonorovskii77,Franks69}. According to this the mean frequency of the signal's $x(t)$ spectrum on a positive semiaxis is defined as a gravity center $\omega_0$  of its energetic spectrum $E(\omega)$:
\begin{equation}\label{omega0}
  \omega_0 =  \int\limits^{\infty}_0{\omega W(\omega) d\omega} = {\int\limits^{\infty}_0{\omega E(\omega) d\omega} \over \int\limits^{\infty}_0{E(\omega) d\omega}},
\end{equation}
where  $E(\omega)=|\dot{S}(\omega)|^2$; $\dot{S}(\omega)=F[x(t)]$  is  a spectrum density of the amplitude of the signal which is limited by the observation interval $[0,T]$. Since the weighting function
\begin{equation}\label{Womega}
  W(\omega) =  { E(\omega)  \over \int\limits^{\infty}_0{E(\omega) d\omega}}
\end{equation}
makes sense of the a posteriori distribution density in the spectrum of the received signal, the estimate \eqref{omega0}  proves to be optimal involving additive noise \cite{Yarlykov85}. This circumstance is justified by the fact that the gravity center of the Doppler frequency spectrum (Fig.~\ref{fig:2}) is determined by the velocity of the geometrical center of the reflecting object whose coordinates and displacement velocity are of interest in the problems of automatic tracking \cite{Yarlykov85}.

However one needs to turn to spectral analysis for an evaluation of spectrum parameters that do not always conform to operative tasks because the spectrum and its characteristics must be calculated after the signal has been received, i.e., beyond the range of observation $[0,T]$.

\section{Fractional differentiation use for estimation of the mean frequency}\label{Fractional}

Real-time frequency measurement poses no problems in the case of monochromatic signal: one just needs to count the number of positive transitions through zero level per unit time (quasi-frequency). This design implies a use of averaging scaling algorithms. We face a problem when the signal of interest widens (Fig.~\ref{fig:2}) true to form in the case of different velocities when the reflecting surface rotates. In this case the velocity estimate at the quasi-frequency value  is not the same as for the true mean \eqref{omega0}); besides the wider the Doppler signal's width, the greater the error.

A rapid evaluation of the spectrum's mean frequency requires the maximum speed of on-line calculations at reception rate of signal counts. Calculation of the energy spectrum $E(\omega)$  and its moments by the soft-hardware-based method using discrete Fourier transformation algorithms (including FFT) \cite{Gonorovskii77} directly through relation \eqref{omega0}) requires a high computational speed and a great amount of computations in frequency domain, because one needs a significant amount of time for the signal counts processing after the expiration of the observation interval $[0,T]$.

\begin{figure}
  \includegraphics[width=1\textwidth]{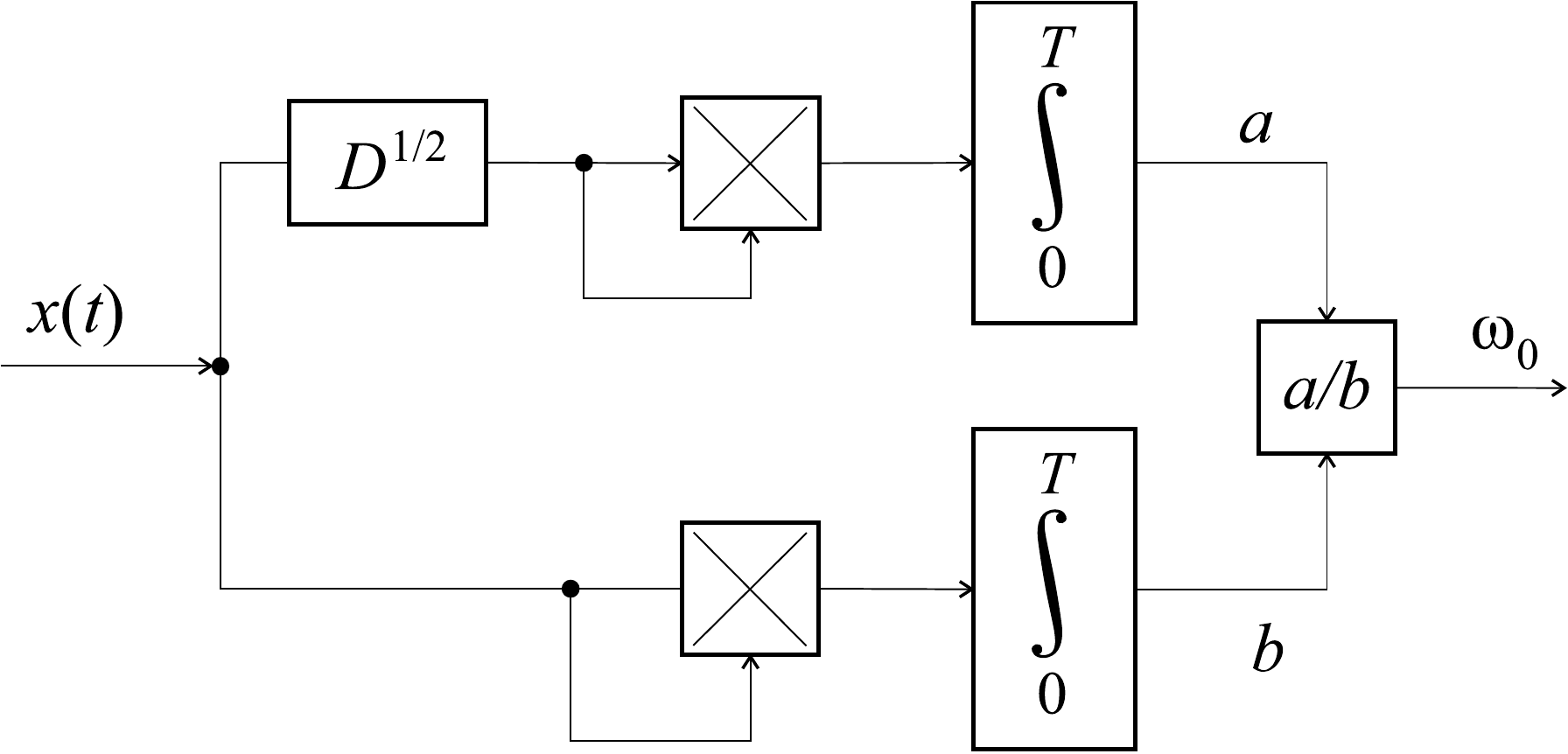}
\caption{The structure of mean frequency $\omega_0$  meter by using the method of fractional differentiation of the Doppler signal $x(t)$.}
\label{fig:3}
\end{figure}

The aim of the proposed approach is an increase in the speed of estimating the Doppler signal mean frequency by computing  $\omega_0$ in the time domain {\it as an on-line signal processing without spectral processing}. Here we resort to an unconventional kind of processing: computation of the signal's fractional derivative as a new part of signal becomes available. The algorithm of fractional differentiation is reduced to realization of a digital filter with special characteristics; this should not cause any major technical problems.

Calculation of square of the  $S(\omega)$  norm in the denominator in \eqref{omega0} can be made in the time domain as on-line processing without spectral processing by transforming the relevant integrals according to Parseval equality \cite{Franks69}
\begin{equation}\label{Somega}
\int\limits^{\infty}_0{|S(\omega)|^2 d\omega} = {1\over 2}\|S(\omega)\|^2 = \pi \|x(t)\|^2 = \pi\int\limits^T_0{x^2(t) dt}.
\end{equation}
One can evaluate the numerator of expression \eqref{omega0} in a similar way which leads to a necessity for fractional differentiation of the signal:
\begin{equation}\label{omegaSomega}
\int\limits^{\infty}_0{\omega |S(\omega)|^2 d\omega} = {1\over 2}\|\sqrt{j\omega}S(\omega)\|^2 = \pi \|D^{1/2} x(t)\|^2 = \pi\int\limits^T_0{|D^{1/2}x(t)|^2 dt},
\end{equation}
where  $D^{1/2}x(t) = \mathbf{F}^{-1}\{\sqrt{j\omega}\mathbf{F}[x(t)]\}$  is a  fractional time derivative of order 1/2 operator determined as convolution of the input signal with the filter pulse response  $h(t) = \mathbf{F}^{-1}\{\sqrt{j\omega}\}$. As a result, expression  \eqref{omega0} can be presented in the form
\begin{equation}\label{omega02}
  \omega_0 =  {\int\limits^{T}_0{ |D^{1/2}x(t)|^2 dt}  \over \int\limits^{T}_0{x^2(t) dt}}.
\end{equation}
Relation \eqref{omega02} shows that the spectrum \eqref{omega0} gravity center estimate can be formed {\it without spectral processing} as a new part of target reflected signal becomes available and can be obtained by the end of observation interval $T$ \cite{Zakharchenko99,Zakharchenko10}.

Fig.~\ref{fig:3} demonstrates the structure of the algorithm of a fast gravity center estimate for the signal's $x(t)$ spectrum in a real-time scale.

The Riemann-Liouville operator of fractional differentiation $D^{\alpha}x(t)$ on positive semiaxis  \cite{Samko93} in the case $\alpha=1/2$  takes the form:
\begin{equation}\label{D12}
D^{1/2} x(t)   =  {1\over\sqrt{\pi}}{ d \over dt} \int\limits^{t}_0{ {x(t^{\prime})\over \sqrt{t-t^{\prime}}} dt^{\prime}}.
\end{equation}
Due to the difference kernel, relation  \eqref{D12} can be treated as Duhamel's integral that couples signals at the input and output of the transversal filter with the pulse response.

\begin{figure}
  \includegraphics[width=1\textwidth]{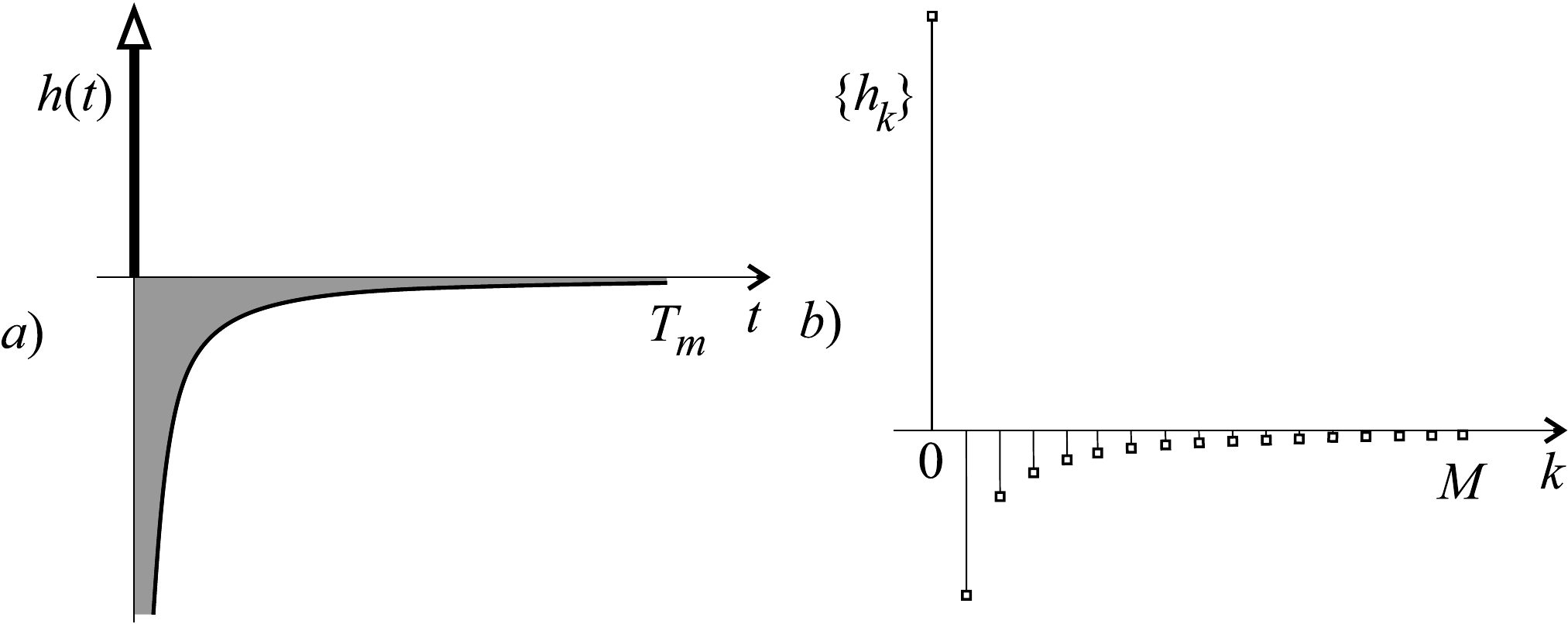}
\caption{The pulse response characteristic $h(t)$ of the fractional differentiating filter of order $1/2$ (a) and its discrete analog (b). The function $h(t)$ is defined to the right of the negative semiaxis and consists of the delta-function defined at zero and continuous monotonically increasing branch below abscissa axis. The function has zero total area under curve (indicated as shaded field).}
\label{fig:4}
\end{figure}

The step response $g(t)$  at $t>0$ represents the response to the Heaviside signal $\sigma(t)$  and can be calculated  substituting  $\sigma(t)$ in equation \eqref{D12}. Taking into account the Heaviside function jump we choose the lower limit of integration as preceded jump by a quantity of $\varepsilon\to 0$. The causality of the step function $g(t<0)=0$  can be satisfied by multiplying the r.h.s. by $\sigma(t)$:
\begin{equation*}\label{g(t)}
\begin{split}
g(t)  & = \sigma(t) \lim_{\varepsilon\to 0} {1\over\sqrt{\pi}}
 { d \over dt} \int\limits^{t}_{-\varepsilon} {\sigma(t^{\prime}) d t^{\prime}\over \sqrt{t-t^{\prime}}} =
\sigma(t) \lim_{\varepsilon\to 0} {2\over\sqrt{\pi}}
 { d \over dt} \int\limits^{\sqrt{t+\varepsilon}}_{0} \sigma(t-\xi^2)d \xi  \\
 & = \lim_{\varepsilon\to 0} {2\sigma(t)\over\sqrt{\pi}}
 { d \over dt} \int\limits^{\sqrt{t+\varepsilon}}_{0} d \xi =
 \lim_{\varepsilon\to 0}  {1\over\sqrt{\pi}} {\sigma(t)\over \sqrt{t+\varepsilon}}.
 \end{split}
 \end{equation*}
The pulse response of the fractional differentiating filter (the eigenfunction of the operator) will have the form (Fig.~\ref{fig:4}):
\begin{equation}\label{h}
h(t)   =   { d \over dt} g(t) = \lim_{\varepsilon\to 0} {1\over\sqrt{\pi}} \left[ {\delta(t)\over \sqrt{t+\varepsilon}}-{\sigma(t)\over 2(t+\varepsilon)^{3/2}}  \right],
\end{equation}
where  $\delta(t)$ is the Dirac delta. It is not difficult to show that this pulse response obeys requirements (4) and can be easily realized. Since the pulse response values (Fig.~\ref{fig:4}a) affect the low-frequency signal's domain at large time, $h(t)$  duration can be limited by the value $T_m\gg 1/\omega_0$  which is relatively insignificant at large cosmic objects velocities.

The value of pulse response time delay $T_m$  in the filter (Fig.~\ref{fig:4}) will dictate the required ram memory for the fractional-differentiating filter operation (Fig.~\ref{fig:4}b: $M=T_m/\Delta T$, where $\Delta T$  is a sampling interval which is much smaller than the ram memory required for the spectral analysis.

\section{Quantitative assessment}\label{Quantitative}

Suppose the NEO has a diameter $\sim 100$ m and rotation  period $\sim 10$ min, then the spectrum width amounts to $\sim 1$ m/s  which corresponds to the Doppler signal's width in the X-band $\sim 50$ Hz. In such a case the mean Doppler frequency equals  $\sim 2$ MHz at velocity 30 km/s. The sampling signal's rate then should be chosen as  $>4$ MHz  ($\Delta T\sim 0.1 \mu$s).

For a better velocity resolution ($\sim 0.01$ m/s or higher \cite{Ostro99}) one has to measure the Doppler frequency with an accuracy of $\sim 0.5$ Hz or higher \cite{Ostro99}. If the Doppler signal duration is $\sim 0.1$ s, this requires processing about   $N\sim 10^6$ signal counts.

The standard spectral analysis with the use of FFT will require  $N \ge 2\cdot 10^7$  multiplication operations. The 10 MFlops  processor will do the task in 2 seconds  whereas the signal lag for the filer of order $M=100$ is only  $T_m=M\Delta t= 100\cdot 10^{-7}$ s $= 10^{-5}$ s. Thus the speed gain is about 5.5 orders of magnitude. The time expenditures will become comparable if teraflops computers are used for  FFT computations which is neither economically advantageous nor technically easily feasible in onboard systems.

According to the Cramer-Rao bound \cite{Cook67}, to achieve such accuracy one has to maintain an excess of signal over noise of at least 30 dB. Since the terminal phase of  target approach when homing characterizes the small distance $R$ which further shortens such an excess maintenance is less of  a problem.

\section{Conclusion}\label{Conclusion}

Thus, the use of fractional differentiation of the Doppler signal allows an exact estimate of velocity of the oversized cosmic objects practically immediately after the arrival of the reflected signal. The proposed algorithm is substantially simpler and more economical than the classical one using the spectral analysis of the Doppler signal.

The authors hope that the suggested algorithm makes the prospect of arranging an asteroid-comet threats protection system more feasible.

The process of real-time estimate of the mean frequency of the Doppler signals by utilizing the differentiation operation is covered by an RF patent \cite{ZakharchenkoPatent}.

%\begin{acknowledgements}
\bigskip\medskip\noindent {\bf Acknowledgements}
\medskip

We thank Vitaly Korolev for assistance in vector graphics drawing and Victor and Vladimir Levi for translation.
%\end{acknowledgements}

%% The Appendices part is started with the command \appendix;
%% appendix sections are then done as normal sections
%% \appendix

%% \section{}
%% \label{}

%% References
%%
%% Following citation commands can be used in the body text:
%% Usage of \citetis as follows:
%%   \cite{key}         ==>>  [#]
%%   \cite[chap. 2]{key} ==>> [#, chap. 2]
%%

%% References with bibTeX database:

%\bibliographystyle{elsarticle-num}
%\bibliography{<your-bib-database>}

%% Authors are advised to submit their bibtex database files. They are
%% requested to list a bibtex style file in the manuscript if they do
%% not want to use elsarticle-num.bst.

%% References without bibTeX database:

\end{document}